# Enhancement in Hydrogen Storage Capacities of Light Metal Functionalized Boron–Graphdiyne Nanosheets

Tanveer Hussain[1,*], Bohayra Mortazavi[2], Hyeonhu Bae[3], Timon Rabczuk[2], Hoonkyung Lee[3] and Amir Karton[1]

[1]School of Molecular Sciences, The University of Western Australia, Perth, WA 6009, Australia

[2]Institute of Structural Mechanics, Bauhaus-Universität Weimar, Marienstr, 15, D-99423, Weimar, Germany

[3]Department of Physics, Konkuk University, Seoul 05029, Republic of Korea

Corresponding Author: tanveer.hussain@uwa.edu.au

## Abstract

The recent experimental synthesis of the two-dimensional (2D) boron-graphdiyne (BGDY) nanosheet has motivated us to investigate its structural, electronic, and energy storage properties. BGDY is a particularly attractive candidate for this purpose due to uniformly distributed pores which can bind the light-metal atoms. Our DFT calculations reveal that BGDY can accommodate multiple light-metal dopants (Li, Na, K, Ca) with significantly high binding energies. The stabilities of metal functionalized BGDY monolayers have been confirmed through *ab initio* molecular dynamics simulations. Furthermore, significant charge-transfer between the dopants and BGDY sheet renders the metal with a significant positive charge, which is a prerequisite for adsorbing hydrogen ($H_2$) molecules with appropriate binding energies. This results in exceptionally high $H_2$ storage capacities of 14.29, 11.11, 9.10 and 8.99 wt% for the Li, Na, K and Ca dopants, respectively. These $H_2$ storage capacities are much higher than many 2D materials such as graphene, graphane, graphdiyne, graphyne, $C_2N$, silicene, and phosphorene. Average $H_2$ adsorption energies for all the studied systems fall within an ideal window of 0.17-0.40 eV/$H_2$. We have also performed thermodynamic analysis to study the adsorption/desorption behavior of $H_2$, which confirms that desorption of the $H_2$ molecules occurs at practical conditions of pressure and temperature.

**Keywords:** 2D Materials, Material design, Hydrogen storage, $H_2$ adsorption, $H_2$ desorption

## 1. Introduction

Due to the continuous increase in global energy utilization and the effects of fossil fuel consumption on the climate, the development of sustainable and renewableenergy supplies is a matter of urgency.Hydrogen ($H_2$) is considered an ideal alternative to the depleting fossil fuels due to attractive properties such as high energy densitywith no harmful effects on the environment. However, an efficient way of storing $H_2$is the main obstacle towards its realization as a green and clean energy carrier.[1-4]Among different storage routes, 2D material-based $H_2$ storage seems to be the most sustainable option provided a suitable material is available, which meets the criteria proposed by US Department of Energy.

Carbon nanostructures (CNs) is a family of materials, which has been studied extensively for $H_2$ storage applications in different morphologies, such as carbon nanotubes, graphene, graphane, graphdiyne, graphyne and many others.[5-14]However, the limitation of CNs, in their pristine form, is the weak binding with $H_2$ molecules, which restricts their applications as efficient $H_2$ storage materials due to small $H_2$uptake orlow operating temperatures.[15] Thus, the binding energies of $H_2$ with the host CNs must be enhanced forstorage at ambient conditions. Several techniques have been employed for improving $H_2$-CNs binding, e.g., spillover effect, defects formation in host materials, application of eclectic fields, inducing charges, and metal functionalization.[16-20]

Functionalization of CNs and other nanostructures by introducing various elementsfor $H_2$ storagehas been comprehensively studied. Zhou and Szpunar investigated the $H_2$ storage properties of graphene sheets dopedwith Pd nanoclusters. They found that Pd clusters of 5–45nm uniformly distributed over the graphene nanosheet attain $H_2$ storage capacity of 6.7 and 8.7 wt% atpressures of 50 and 60 bar, respectively.[21]Zhou *et al.* synthesized Ni/graphene compositesin which Ni clusters of 10 nm are uniformlydispersed over the graphene surface and studied their $H_2$ storage capacities. At ambient conditions an $H_2$ storage capacity of 0.1 wt% was achieved, andit could be increased to 1.2 wt% at a pressure of60 bar.[22]

Theoretical studies of metal functionalized CNs other than graphene for $H_2$ storage have also been carried out. Liu *et al.* used first principles calculations based on density functional theory (DFT) to study the structural, and $H_2$ storage properties of Mg doped γ-graphyne sheet and obtained a storage capacity of 10.6 wt%.[23] It was reported that the application of electric field would restrict the aggression of Mg dopants and the adsorbed $H_2$ molecules bind with an energy range of 0.28 eV/$H_2$, which falls within the desired window. Pan *et al.* used DFT calculations to study the $H_2$ storage properties of various CNs such as $C_{40}$, $C_{41}$, $C_{63}$, $C_{64}$ and $C_{65}$ doped with Ca at various concentrations. They concluded that Ca binds strongly with the CNs and anchor multiple of $H_2$ molecules attaining a high $H_2$ storage capacity of up to 8.6 wt%.[24] Another DFT study by Mohajeri and Shahsavar investigated metal functionalization of a graphyne monolayer under nitrogen and sulphur co-doping. They reported high $H_2$ storage capacities of 9.0 and 9.3 wt% for Li and Na metal doping, respectively.[25]

The above studies demonstrate that metal functionalization of CNs can play a vital role in improving $H_2$ binding with the host material. A very recent addition to the family of two-dimensional CNs is the boron-graphdiyne (BGDY) nanosheet, which has been synthesized via a bottom-up synesthetic approach.[26] Among several attractive properties of BGDY such as enhanced optical, thermal stability, mechanical response, and thermal conductivity, the presence of boron centres uniformly distributed in a carbon network creates additional binding cites for metal centres. Mortazavi *et al.* studied the structural, electronic, thermal, mechanical, optical and metal storage properties of BGDY by means of DFT coupled with molecular dynamics simulations.[27] Under the application of mechanical strain, it was concluded that this porous 2D monolayer preserves superstretchability. The authors further studied the application of BGDY as high capacity anode material for Li, Na and Ca ion batteries. Motivated by the enhanced metal storage properties of BGDY, we have employed spin-polarized DFT-D3 calculations to study its potential use as a high capacity $H_2$ storage material. We have considered both alkali (Li, Na, K) and alkaline earth metal (Mg, Ca) dopants uniformly distributed over the BGDY monolayer, and studied their structural, electronic and $H_2$ storage capacities. Our simulations revealed an optimum level of metal doping concentration for efficient $H_2$ storage.

## 2. Computational details

We have carried out spin-polarized periodic boundary condition DFT calculations using the VASP code.[28,29]In these calculations, the generalized gradient approximation (GGA) PBE exchange-correlation functional has been used.[30]We have employed projector-augmented wave (PAW) method to deal with theion–electron interactions.[31]Empirical van der Waals correctionshave been included using the D3 dispersion correction of Grimme *et al*.[32]The Brillouin zone (BZ) has been sampled by Monkhorst–Pack scheme with a mesh size of 3×3×1 for geometry optimization and 7×7×1 for obtaining density of states.[33]A vacuum space of 20 Å has been inserted, which is large enough to evade the possible interactions between periodic images along the z-axis. In the geometry optimizations, the convergence criteria for the total energies and forces have been set at $10^{-5}$ eV and 0.01 eV/Å, respectively. Binding of metal dopants to the BGDY monolayer involve charge transfer mechanism, which has been studied by mans ofBader charge analysis.[34]Binding energies per dopant ($E_b$) of metal adatoms on BGDY are calculated using the following equation:

$$E_b = \{E(GDY+nX) - E(BGDY) - nE(X)\}/\underline{n} \quad (1)$$

where X=Li, Na, K, Mg, Ca, and $n$=1–4. In this equation, the first, second and third terms represent the total energies of BGDY bonded with $n$metal dopants, pristine BGDY monolayer and metal dopants, respectively.

## 3. Results and discussion

The optimized structure of BGDY is shown in Figure 1. The BGDY monolayer used in this study has 14 atoms ($C_{12}B_2$) consisting of two types of C–C bond lengths of 1.23Å and 1.35Å, and a B–C bond length of 1.51Å. The calculated lattice constant of 11.85Å and the bond lengths mentioned above agree well with the previous study which used the more reliable hybrid GGA functional HSE06.[27]In pristine form, BGDY preserves a semiconducting behavior with aenergy gap ($E_g$) of 0.485 eV, as evident from a total and partial density of states plots (Figure 1(c)). However, this $E_g$ value is underestimated due to the well-knowninability of the PBE functional to calculate the exact $E_g$. A more accurate $E_g$ value has been calculated by Bohayra *et al.*, using the HSE06 functional they obtained a value of 1.15 eV.[27]As the mechanical, thermal and dynamic stability of the BGDY monolayer has already beenstudiedin Reference **[27],** we start by investigating its metal doping capacities.

Like most of the CNs, pristine BGDY barely binds $H_2$, thus metal dopants (e.g., Li, Na, K, Mg, Ca) have to be introduced to enhance the BGDY-$H_2$ binding energies. The binding energies of these dopants over the BGDY monolayer should exceed theircorresponding cohesive energies ($E_c$) to ensure a reversible doping mechanism. On the other hand, when $E_c$ exceeds $E_b$ clustering of metal dopantswould be a more likely outcome rather than binding to a BGDY. The selection of alkali and alkaline metal dopants has been based on the fact that these light elements, having lower cohesive energies, would make uniform scattering over the BGDY monolayer. We have considered all the possible binding sites on BGDY monolayer for metal doping and compare the $E_b$ values.The lowest $E_b$ configurations yield $E_b$ values of -2.95, -2.51, -2.91, -1.13 and -3.04 eV for Li, Na, K, Mg and Ca, respectively. With the exception of Mg (-1.51 eV), these $E_b$ values are much higher than the corresponding cohesive energies of Li (-1.63 eV), Na (-1.11 eV), K (-0.93 eV) and Ca (-1.84 eV).[35, 36]Moreover,our calculated $E_b$ values for the studied dopants are higher than those of graphenylene, phosphorene, stanene, siligraphene ($SiC_7$) and graphdiyne nanosheets.[5,20,37-39]This indicates the potential of BGDY as a promising metal anchoring material.

Large surface to volume ratio due to its 2D nature and big pore size enables BGDY to accommodate more metal dopants. This would assist in achieving more active sites for the incoming $H_2$ molecules to be adsorbedon BGDY monolayers, which would result into a large $H_2$ gravimetric density. However, the introduction of metal dopants on BGDY will be associated with the transfer of charge from the former to the later. Thus, electrostatic repulsionbetween the cationic dopants could be a concern in achieving high metal doping concentration and consequently high $H_2$ storage capacity. Maintaining a reasonably high distance between the two adjacent metal dopants can solve this problem. For further metal adsorption, the BGDY doped with single metal dopant has been considered as initial structure. The second dopant is introduced on all possible binding sites availableon BGDY monolayer preserving a sufficientlylarge distance from the existing metal atom. The lowest energy structures among the considered configurations are associated with $E_b$ values of -2.62, -2.10, - 2.45 and -2.61 eV, for Li, Na, K and Ca, respectively. Upon the introduction of second dopant, the reduction in $E_b$ values are 11.19, 16.33, 15.80 and 14.14% for Li, Na, K and Ca, respectively. Despite the decrease in $E_b$, we still obtain $E_b > E_c$, which indicatesthat the dopants dispersion over BGDY is preferred instead of metal clustering. We kept on adding more metal dopants as long as their $E_b$ is exceeding the $E_c$ values. The optimized structures of BGDY monolayers with maximum metal dopants considered in this study are shown in Figure 2. For the third dopant, BGDY monolayer pre-adsorbed with two dopants and for the fourth dopant, BGDYpre-adsorbed with three dopants served as an initial structure. It is important to mention here that the minimum dopant-dopant distance at the highest doping concentrations for Li, Na, K and Ca has been found as 4.45, 5.16, 5.20 and 4.86 Å, respectively. Complete results for the structural parameters of all the systems at various doping concentrations are given in table 1. The $E_b$ values in the case of four dopants for all the metals are still higher than their corresponding $E_c$as shown in Figure 3 (a).

In addition to the energetic stability, that is $E_b > E_c$, we have confirmed the thermal stabilities of BGDY monolayers loaded with four metal dopants of Li, Na, K and Ca through *abinition* molecular dynamics simulations (AIMD). This is done by employing Nose-thermostat algorithm at 300 K for 6ps with a time step of 1fs. All the doped systems remained stable without significant structural deformation, which is evident from small variations in their energies (for further details see Figure S1of the Supplementary Information). This ensures the thermal stabilities of metal-doped BGDY monolayers.

As mentioned above, the metal-BGDY interactions involve charge transfer, we have employed Bader analysis to estimate the amount of charge transfer between the dopants to the BGDY monolayers. At a maximum doping concentration, that is BGDY-4X (X=Li, Na, K, Ca) each Li, Na, K and Ca dopants have transferred an average charge of 0.98, 0.68, 0.60 and 1.09 *e* to the monolayer, respectively. This implies that BGDY monolayers attain significantly high negative charges, leaving the dopants in cationic form with reasonably high partial positive charges. Bulk share of these donated charges has been captured by the C and B atoms of the BGDY monolayer, which are in close vicinity to the dopants. Depletion and accumulation of electronic charges has been calculated by the following relation and shown in Figure 4.

$$\Delta \rho = \rho(\text{BGDY: X}) - \rho(\text{BGDY}) - \rho(\text{X}) \quad (2)$$

Here the first, second and third terms represent the charge densities of doped BGDY, pristine BGDY, and the metal dopant, respectively.

Transfer of charges from the dopants would change the electronic properties of BGDY monolayers, which have been studied by density of states plots. Figure 5shows the partial density of states(PDOS) plot of BGDY doped with 4Li atoms. One can see a transition from semiconducting to metallic BDGY upon the introduction of Li adatoms. The contribution appearing at the Fermi level ($E_f$) is from Li(s), which clearly overlaps with B(p) and C(p). This Li dopant is the one, which is bonded to BGDY monolayer in the vicinity of both B and C atoms. As indicated by in the Bader charge analysis that each Li dopant donates an average of 0.98$e$to BGDY, thus the distinct peaks appearing between -3.70 to -3.50 eV of the left, and then 0 to 2.0 eV on the right side of $E_f$ corresponds to Li(s) contributions to BGDY monolayer.

Similar to Li, the other dopants, Na, K and Ca, cause semiconducting to metallic transitions as shown in Figures S2–S4 (supplementary information). For Na doping, the distinguishing peaks of Na(s) appear at -4.2, -3.8 and -0.92 eV on the left and at 0.20, 0.40 and 1.20 eV on the right of $E_f$.For K and Ca doping, the valence bands are mainly dominated by C(p) and B(p), with small contributions from the$s$ orbitals of K and Ca. However, at the top of $E_f$the hybridization of K(s) and Ca(s) with those of C(p) and B(p) are more dominated as compared to the Li and Na cases, which is evident in Figures S3 and S4, respectively.

So far we have discussed the structural, thermal stabilities, charge transfer and electronic properties, we will now move on to investigate the adsorption of $H_2$on metal-functionalized BGDY monolayers. Each system has four metal dopants, which are positively charged and capable of binding $H_2$ molecules through electrostatic as well as van der Waals interactions. Each of the positively charged ions($Li^{\delta+}$, $Na^{\delta+}$, $K^{\delta+}$, $Ca^{\delta+}$) generates a local electric field though which it polarizes the incoming $H_2$ molecules and thus induces a local negative charge on $H_2$. Induction of a charge on $H_2$resultsin a binding strength, which is much stronger than that of pristine BGDY-$H_2$ binding and can be calculated by the following equation:

$$E_b(H_2) = \{E(BGDY@4X:nH_2) - E(BGDY@4X) - nE(H_2)\}/n \quad (3)$$

In this equation, $E_b(H_2)$ is the binding energies of $H_2$ on metal functionalized BGDY, andthe first, second and third terms on the right-hand-side are the total energies of BGDY@4X bonded with n $H_2$ molecules, pristine BGDY@4X and $H_2$, respectively(where X=Li, Na, K and Ca and *n*=1–4).

To avail all the active metal binding sites, we have introduced $H_2$ molecules on each dopant of the functionalized BGDY monolayers in a stepwise mode. That is, in the first step four $H_2$ molecules have been inserted on BGDY@4X and the systems are allowed to relax completely. Based on our previous experience,[12] a vertical adsorption of $H_2$ to the metal dopant is more favorable.In the next step, we introduced a second $H_2$ molecule on each dopant of the optimized BGDY@4X carrying 4$H_2$ molecules. For further $H_2$ adsorption, a suitable distance of $H_2$ with the metal dopants as well as among charged induced $H_2$ molecules should be maintained to avoid the steric and electrostatic repulsion.Overall,each metal dopant X of BGDY@4Xis bondedto two $H_2$ molecules, thus a total of 8$H_2$ molecules have been adsorbed. This process is repeated until the systems reach saturation, that is the point where further $H_2$molecules would be repelled upon optimization.

It is important to mention here that the binding energies per $H_2$molecule (eq.3)is compared with the desired range of 0.15–0.60 eV/$H_2$each time an $H_2$ is added to the system. We have concluded that each dopant can accommodate a maximum of 4$H_2$ molecules, which means that a total of 16$H_2$ could be adorned on BGDY@4X with $H_2$ binding energies within the accepted range. The van der Waals corrected DFT $E_b(H_2)$ for all the systems with different $H_2$ coverage are given in Figure 3 (b). One can clearly notice a decreasing trend in the $E_b(H_2)$ values upon the increasing $H_2$ densities in Figure 3 (b). In case of BGDY@4Li, the binding energies vary from -0.297 to -0.170 eV per $H_2$ from minimum to maximum $H_2$ coverage. For the other dopants we obtain the following ranges of $E_b(H_2)$ per $H_2$from minimum to maximum $H_2$ coverage:-0.343 to -0.186 (BGDY@4Na), -0.389 to -0.187 (BGDY@4K) and -0.408 to -0.206 (BGDY@4Ca) eV.

At a maximum $H_2$ coverage the $H_2$ storage capacities of 14.29, 11.11, 9.10 and 8.99 wt% have been achieved for BGDY@4Li, BGDY@4Na, BGDY@4K, BGDY@4Ca, respectively. These storage capacities are superior to those of other metal-doped monolayers such as graphene, h-BN, $C_2N$, GDY, $MoS_2$, phosphoreneand silicene as shown in table 1. [5,10,40-44] Optimized structures of metal-functionalized BGDY under hydrogenation are given in figure 6.

For an ideal $H_2$ storage material, in addition to a highcontent of $H_2$ adsorption, desorption of $H_2$ molecules at feasible operating conditions is of greatimportance. Thus, it is essential to investigate the desorption capacities of $H_2$ under practical conditions of pressure and temperature. For this purpose, we have employed the thermodynamic analysis of the adsorption of $H_2$ on light metal-doped BGDY monolayers for a given temperature and pressure to understand thermodynamics of $H_2$. The number of $H_2$ molecules can be calculated from the following formula.[45, 46]

$$N_{H_2}(P,T) = N_0 \frac{\sum n g_n e^{n(\mu-\varepsilon_n)/k_B T}}{\sum g_n e^{n(\mu-\varepsilon_n)/k_B T}}, \qquad (4)$$

where $N_0$ denotes the maximum number of adsorbed $H_2$ molecules, $n$ indicates the number of $H_2$ molecules, $\mu$ denotes the chemical potential of the $H_2$ gas, and $\varepsilon_n$ and $g_n$ denote the adsorption energy per $H_2$ molecule and degeneracy of the configuration, respectively. The chemical potential of $H_2$ gas was used for the experimental values.[47] Using the Eq. (4), the number of $H_2$ molecules was calculated for BGDY@4X as shown in Figure 7(a). The systems GBDY@4Li and BGDY@4Narequire temperature lower than the others because the $E_b$ values of $H_2$ on Li and Na are smaller than that on K and Ca. In addition, the change of the number of $H_2$ molecules is not linearly or parabolic, that is, there are multiple flatregionsas the temperature decreases. This is ascribed to the fact that the adsorption is generated from the multiple binding energy of $H_2$ according to the number of adsorbed $H_2$ molecules. On the other hand, at higher pressure the adsorption occurs at a higher temperature as shown in Figure 7(a) because the chemical potential of the $H_2$ gas increases as the pressure increases.[48] Thus, at constant temperature, $H_2$ can be released as the pressure decreases (Figure 7(b)), which is an isothermal adsorption-desorption process.

From figure 7 (a, b) it can be clearly seen that the all the systems can contain a large number of $H_2$ molecules at practical operating conditions of pressure and temperature, which further confirms the effectiveness of light metal doped BGDY monolayers for high capacity H$_2$ storage applications.

| System | Type of study | Dopants | Hydrogen Storage Capacity (wt%) | Reference |
|---|---|---|---|---|
| Graphene | Experimental | Pd | 6.7 | 21 |
| Graphene | Experimental | Ni | 0.1-1.2 | 22 |
| γ-graphyne | Theoretical (DFT) | Mg | 10.6 | 23 |
| Graphyne | Theoretical (DFT) | Li, Na | 9-9.3 | 25 |
| Graphdiyne | Theoretical (DFT) | Li, Na, K, Ca, Sc, Ti | 4.91-6.5 | 5 |
| Graphene | Theoretical (DFT) | Cu | 4.23 | 4 |
| C$_2$N Sheets | Theoretical (DFT) | Li | 10 | 41 |
| MoS$_2$ | Theoretical (DFT) | Li | 4.4 | 42 |
| Phosphorene | Theoretical (DFT) | Li | 5.3 | 43 |

Table 1: Hydrogen storage capacity values from different 2D materials.

## 4. Conclusion:

In summary, we have employed spin-polarized DFT-D3 calculations to study the structural, electronic, charge transfer and H$_2$ storage properties of recently synthesized BGDY monolayers. Our simulations reveal that BGDY binds light metal dopants such as Li, Na, K and Ca with binding energiesof -2.95, -2.51, -2.91, and -3.04 eV, respectively. Even at high metal doping concentrations, the metal bindings are significantly higher than the dopant's experimental cohesive energies, thus achieving a uniform metal distribution over the BGDY surface without clustering. In addition, we have alsoperformed AIMD simulationsat 400K to confirm the thermal stabilities of metal-doped BGDY monolayers. Bader charge analysis reveals that an average of 0.98, 0.68, 0.60 and 1.07$e$have been transferred from Li, Na, K and Ca, respectively to the BGDY sheet, which means the dopants carry a significant positive charge. We find that each BGDY unit cell can accommodate four metal dopants of each type (Li, Na, K or Ca).These metal dopants can adsorb a maximum of 16 $H_2$ molecules with an average $E_b$ ($H_2$) within an ideal range of $E_b = 0.17–0.40$ eV per $H_2$ molecule. Thus, the doped BGDY monolayer attains $H_2$ storage capacities which are much higher than other 2D systems reported in the literature such as graphene, graphane, graphdiyne, graphyne, $C_2N$, g-$C_3N_4$, h-BN, MXene, silicene, phosphorene. In order to design efficient $H_2$ storage systems, which could operate under practical operating conditionsof pressure and temperature, we have studied the thermodynamic properties of metal-doped BGDY monolayers. Our comprehensive thermodynamic analysis reveals that the $H_2$ molecules adsorbed on the doped BGDY layer could be desorbed under practical conditions of pressure and temperature. Thus, metal functionalized BGDY is a promising material for efficient, reversible and high capacity $H_2$ storage under ambient conditions applications.

**Acknowledgments:**

This research was undertaken with the assistance of resources from the National Computational Infrastructure (NCI), which is supported by the Australian Government. AK acknowledges an Australian Research Council (ARC) Future Fellowship (FT170100373). HL was supported by the Basic Science Research Program (Grant No.KRF-2018R1D1A1B07046751) through the National Research Foundation (NRF) of Korea, funded by the Ministry of Education, Science and Technology. BM and TR acknowledge the financial support by European Research Council for COMBAT project (Grant number: 615132).BM also acknowledges *Cluster of Excellence PhoenixD (Photonics, Optics, and Engineering–Innovation Across Disciplines),* Leibniz Universität Hannover, *Hannover, Germany.*

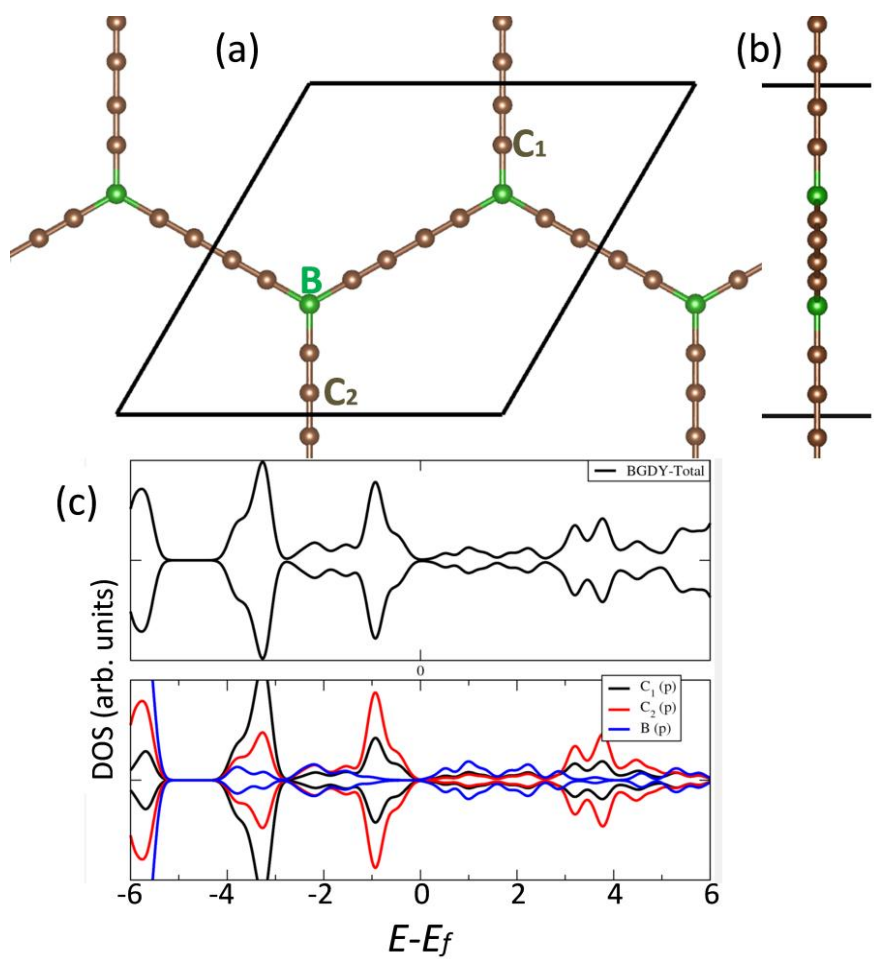

Figure 1. Top (a) and side (b) view of ground state configuration of BGDY monolayer. Brown and green balls represent C and B atoms, respectively. Total and partial densities of states plots are also given in (c).

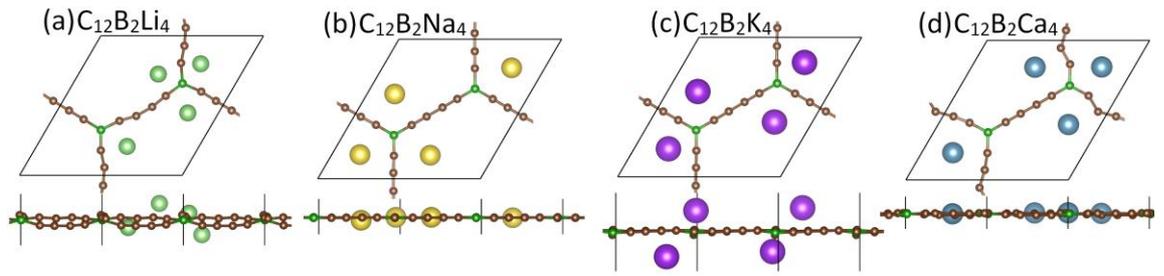

Figure 2. Top and side views of the optimized structures of (a) BGDY@4Li, (b) BGDY@4Na, (c) BGDY@4K and (d) BGDY@4Ca. Brown, dark green, light green, yellow, purple and blue atoms represent C, B, Li, Na, K and Ca, respectively.

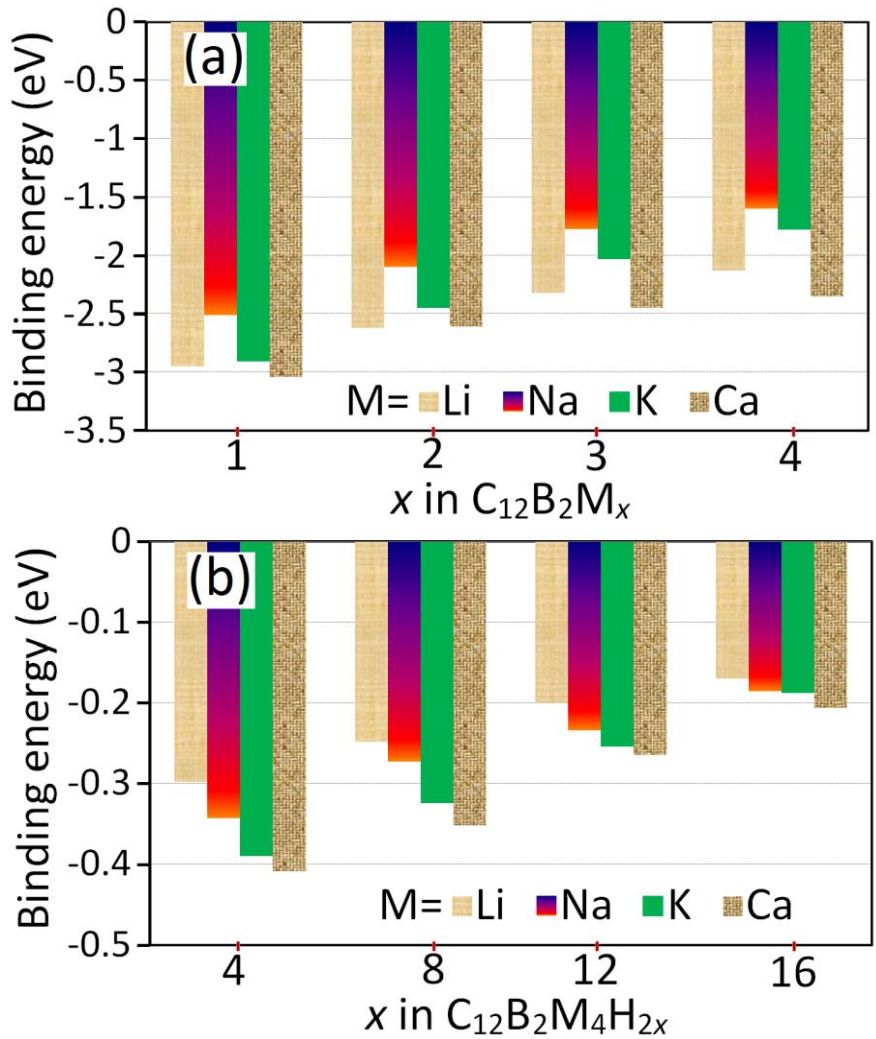

Figure 3. (a) Binding energies ($E_b$) of metal dopants over BGDY monolayer at different doping concentrations. (b) Average $H_2$ adsorption energies (per $H_2$ molecule) on BGDY loaded with four metal dopants.

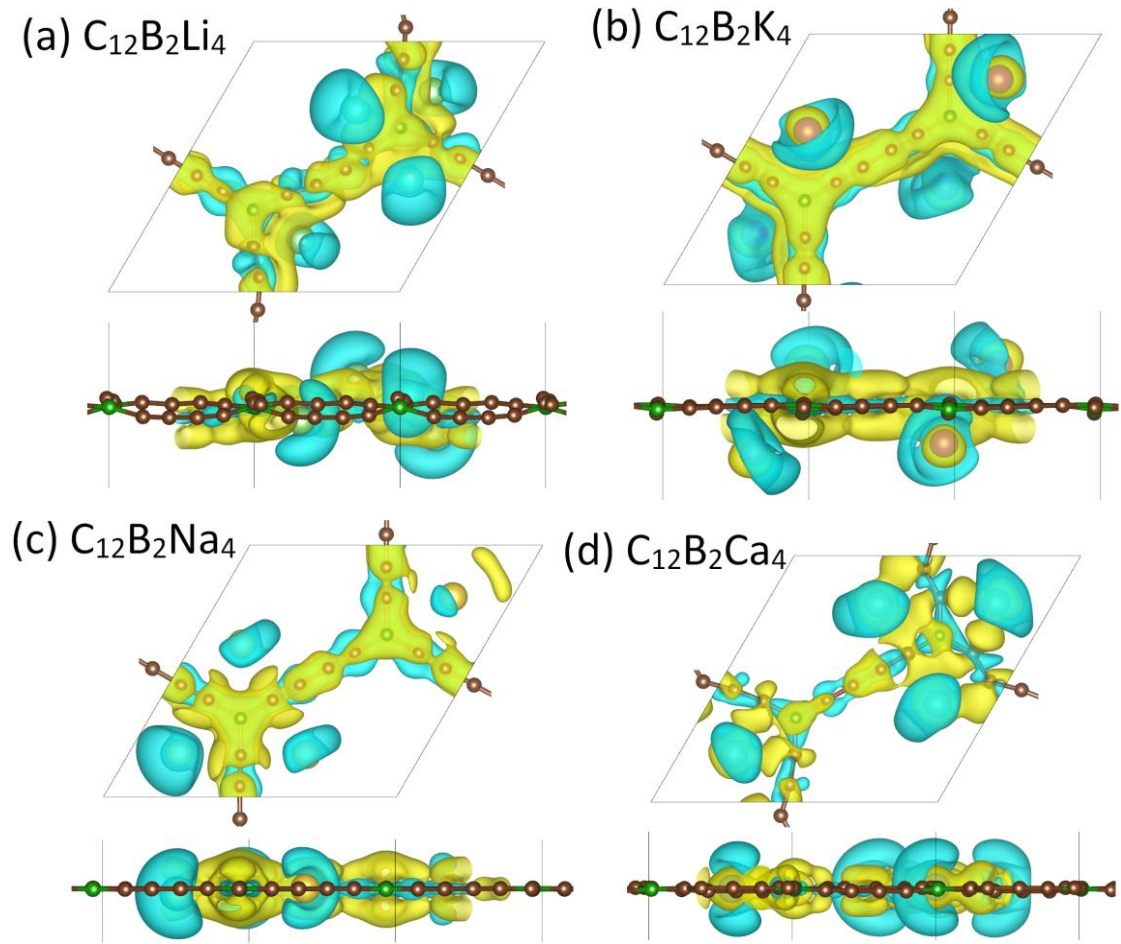

Figure 4. Isosurface charge densities of Li, Na, K and Ca doped BGDY monolayers at maximum doping concentrations with isovalue of 0.01 e/Å$^3$. Yellow and cyan surfaces indicate the accumulation and depletion of charges, respectively.

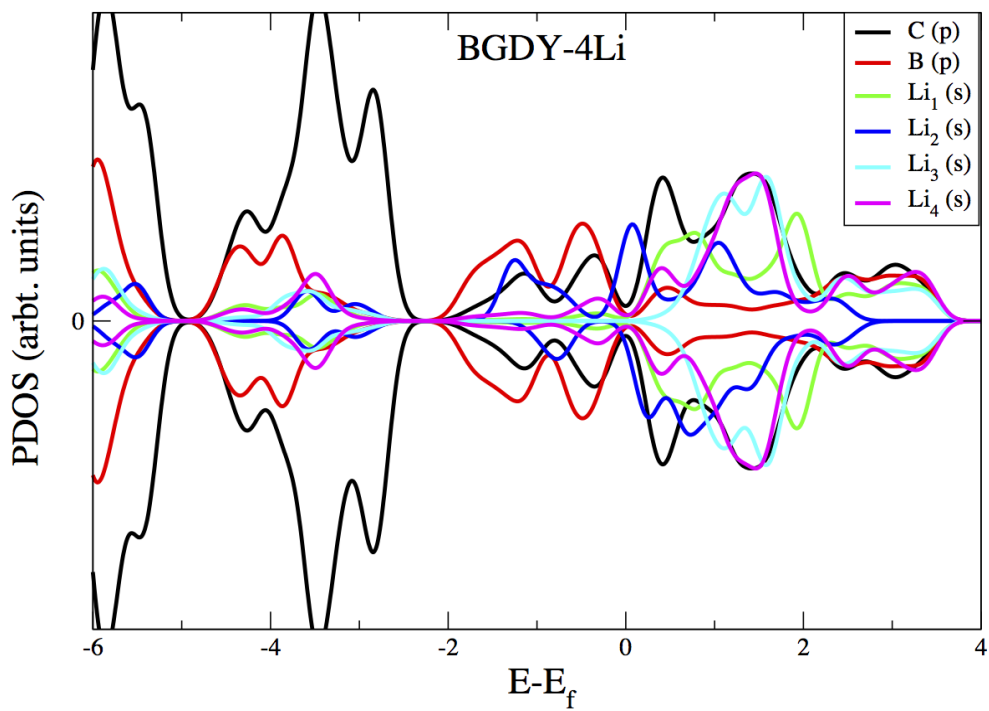

Figure 5. PDOS plots of BGDY-4Li monolayer. The Fermi level is set to zero.

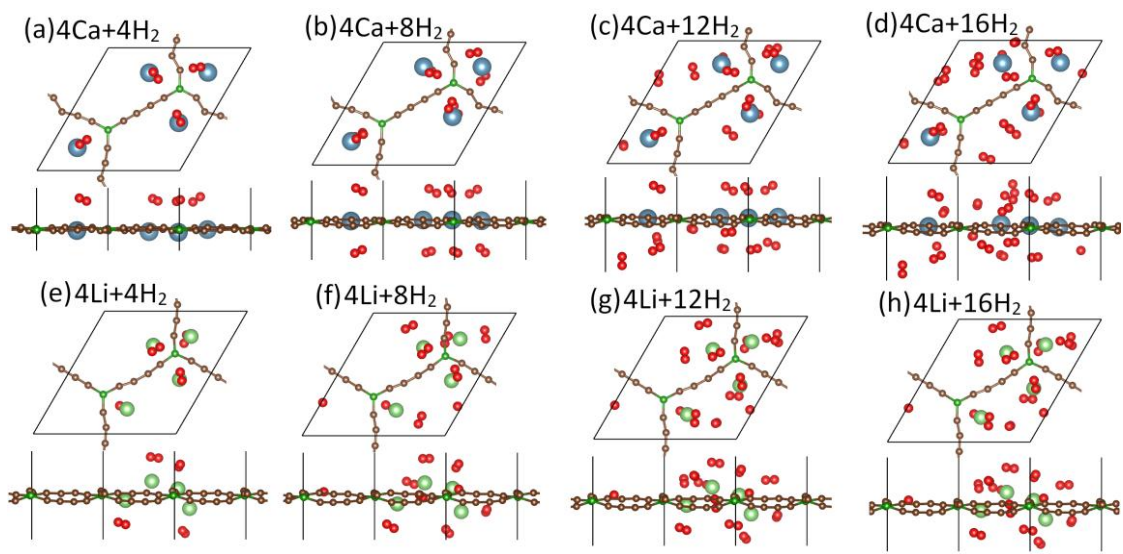

Figure 6. Top and side views of the optimized structures of (a-d) BGDY@4Ca and (e-h) BGDY@4Li loaded with $4H_2$, $8H_2$, $12H_2$ and $16H_2$, respectively. Brown, dark green, blue, light green, and red atoms represent C, B, Ca, Li and H, respectively.

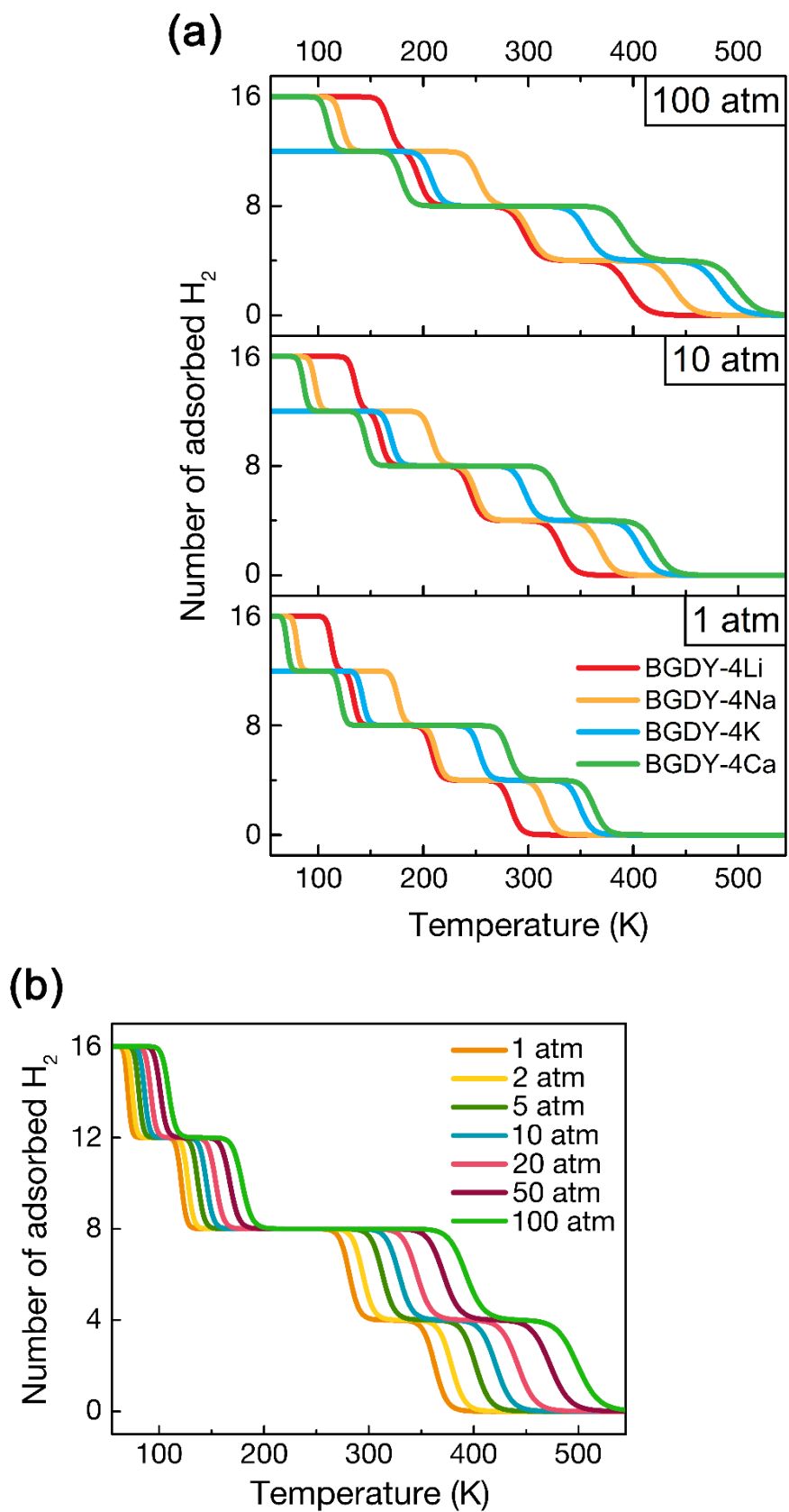

Figure 7. The thermodynamics of H$_2$ adsorption on alkali-metal-doped BGDY. (a) The adsorption of H$_2$ molecules on alkali-metal-doped BGDYs, as a function of the temperature and the pressure of H$_2$ gas. (b) Adsorption-desorption process by changing pressure of H$_2$, for BGDY-4Ca.

| System | E$_b$ per X (eV) | E$_b$/E$_c$ | D$_1$ (Å) X-BGDY distance | D$_2$ (Å) X-X distance | Charge transferred Q (é) |
|---|---|---|---|---|---|
| BGDY-Li | -2.95 | 1.81 | 2.10 | 11.84 | 0.992 |
| BGDY-2Li | -2.62 | 1.60 | 2.16 | 6.02 | 0.991 |
| BGDY-3Li | -2.32 | 1.42 | 2.20 | 4.83 | 0.987 |
| BGDY-4Li | -2.13 | 1.29 | 2.29 | 4.45 | 0.980 |
| BGDY-Na | -2.51 | 2.61 | 2.39 | 11.84 | 0.992 |
| BGDY-2Na | -2.10 | 1.89 | 2.43 | 5.93 | 0.991 |
| BGDY-3Na | -1.77 | 1.59 | 2.53 | 5.88 | 0.747 |
| BGDY-4Na | -1.60 | 1.44 | 2.58 | 5.16 | 0.680 |
| BGDY-K | -2.91 | 3.13 | 2.73 | 11.84 | 0.914 |
| BGDY-2K | -2.45 | 2.63 | 2.80 | 6.79 | 0.893 |
| BGDY-3K | -2.03 | 2.18 | 2.81 | 5.81 | 0.727 |
| BGDY-4K | -1.78 | 1.91 | 2.85 | 5.20 | 0.600 |
| BGDY-Ca | -3.04 | 1.65 | 2.39 | 11.84 | 1.459 |
| BGDY-2Ca | -2.61 | 1.41 | 2.46 | 8.71 | 1.139 |
| BGDY-3Ca | -2.44 | 1.32 | 2.40 | 5.45 | 1.107 |
| BGDY-4Ca | -2.35 | 1.27 | 2.48 | 4.86 | 1.070 |

Table 1. Binding energies (E$_b$), binding to cohesive energies ratio (E$_b$/E$_c$), average dopants-BGDY distance (D$_1$), average dopant-dopant distance (D$_2$) and average charge transferred from dopants to BGDY monolayers (Q). X=Li, Na, K, Ca